\documentclass[pra,showkeys,twocolumn,aps]{revtex4-1}
\usepackage{graphicx}

\begin{document}
\title{Solid argon as a possible substrate for quasi-freestanding silicene}
\author{S. Sattar$^1$}
\author{R. Hoffmann$^2$}
\author{U. Schwingenschl\"{o}gl$^{1,}$}
\email{udo.schwingenschlogl@kaust.edu.sa, +966(0)544700080}
\affiliation{$^1$PSE Division, KAUST, Thuwal 23955-6900, Kingdom of Saudi Arabia}
\affiliation{$^2$Department of Chemistry and Chemical Biology, Cornell University,
Baker Laboratory, Ithaca NY 14853-1301 (USA)}
\date{\today}


\begin{abstract}
We study the structural and electronic properties of silicene on solid Ar(111) substrate using
ab-initio calculations. We demonstrate that due to weak interaction quasi-freestanding silicene
is realized in this system. The small binding energy of only $-32$ meV per Si atom also
indicates the possibility to separate silicene from the solid Ar(111) substrate. In addition,
a band gap of $11$ meV and a significant splitting of the energy levels due to spin-orbit
coupling are observed.
\end{abstract}

\maketitle

\section{Introduction}
Silicene is based on a two-dimensional honeycomb Si lattice, similar to
graphene, attracting interest by the predicted linear dispersion near the Dirac point
and various potential applications in electronic devices. The structure is not
perfectly two-dimensional, but a finite buckling mixes some $sp^3$ hybridization
into the $sp^2$ states. The weaker Si-Si bonding, as compared to the
C-C bonding in graphene \cite{2dbands}, strongly complicates the synthesis. Still,
silicene has been grown on thin film ZrB$_2$, resulting in electronic properties
that are different from the expectations for freestanding samples \cite{zrb2_1}.
Both the buckling and the electronic properties can be modified by epitaxial strain,
which hints at a strong interaction with the substrate. Silicene
on Ir(111) has been investigated experimentally and theoretically in Ref.\ \cite{ir111_si}.

It has been demonstrated that Si nanoribbons can be grown on Ag(110) substratea
\cite{si_ag1101,si_ag1102} and the electronic structure has been investigated by
angular resolved photoelectron spectroscopy \cite{si_ag1103}.
Epitaxial growth with a highly ordered honeycomb structure
on Ag(111) has been confirmed by scanning tunneling microscopy \cite{ag111_si1}.
The Si nearest neighbor distance of $1.9\pm0.1$ \AA, obtained by line analysis
of the microscopy data, also points to strong interaction with the substrate.
A systematic study of Si superstructures on Ag(111) has been performed in Ref.\
\onlinecite{ag111_si3} using low energy electron diffraction, scanning tunneling
microscopy, and ab-initio calculations. Further results from scanning tunneling
microscopy have been reported in Refs.\ \cite{ag111_si4,ag111_si5,ag111_si6},
confirming that the quasiparticles in silicene behave as massless Dirac fermions.
However, experiments indicate that the Dirac nature is
perturbed by symmetry breaking due to the substrate \cite{ag111_si7}. This argumentation
is supported by ab-initio results, which lack a Dirac dispersion for various stable and
metastable structures of silicene on Ag(111) \cite{ag111_si8, ag111_si9}.

Graphene has been separated from SiC(0001) substrate, on which the binding energy
experimentally amounts to 106 meV \cite{Ebg_SiC}. As Si bonds are usually weaker
than C bonds, to separate silicene from a substrate probably a significantly
smaller energy will be necessary. Hexagonal boron nitride
\cite{grap-hbn} and SiC(0001) \cite{grap-sic1,grap-sic2} are known
substrates for graphene and therefore have been studied also for
silicene by ab-initio calculations, finding that the Dirac cone is
preserved, though slightly doped in the case of hydrogenated SiC(0001)
\cite{si-hbn}. For a superlattice of silicene and hexagonal boron
nitride a binding energy of 57 meV per Si atom has been predicted
theoretically \cite{si-hbn2}. The electronic properties of silicene
on II-VI and III-V semiconducting (111) substrates, including AlAs,
AlP, GaAs, GaP, ZnS, and ZnSe have been investigated in some detail \cite{si-semisub,new1,new2},
finding n-doping on metal terminated and p-doping on non-metal terminated surfaces.

Ar exists as solid at low temperature, with short range, weak, and attractive London
dispersion forces responsible for the molecular bonding \cite{argonforce}.
The temperature-pressure phase diagram has been studied in Refs.\ \cite{young75,macrander77},
demonstrating that a face centered cubic structure is stable below a temperature of 84 K.
Despite various attempts, so far no suitable substrate could be identified such that
the characteristic electronic structure of silicene would not be perturbed dramatically on it
\cite{dimoulas13}. This is probably the consequence of too high binding energies
on the tested substrates. For example, values of 89 and 76/84 meV per atom have been reported
for hexagonal boron nitride and Si/C-terminated SiC(0001), respectively \cite{si-hbn}.
In this context, we study the possibility of utilizing solid Ar(111) as a substrate and
analyze the consequences on the electronic properties of silicene. We will argue that
silicene on solid Ar(111) is quasi-freestanding. In addition, in Refs.\ \cite{new3,new4}
a buffer layer of solid noble gas has been used to deposit metal clusters by soft landing
and subsequent evaporation of the noble gas. A similar approach with solid Ar(111) on top
of the desired substrate can provide access to growth of silicene on essentially any substrate.

\begin{figure}[b]
\includegraphics[width=0.45\textwidth,clip]{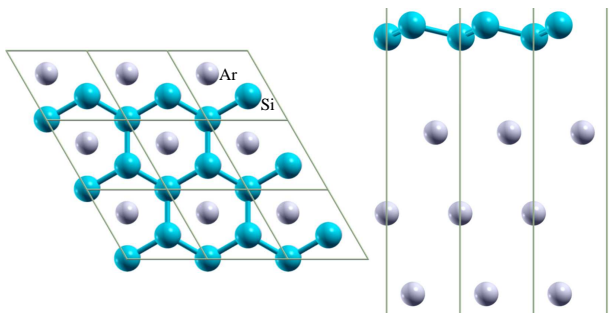}
\caption{\label{fig:struct}Top view (left; along c-axis) and side view
(right; along x-axis) of silicene on Ar(111).} 
\end{figure}

\section{Computational method}

All calculations are performed using density functional theory
in the generalized gradient approximation (Perdew-Burke-Ernzerhof parametrization) and ultrasoft pseudopotentials,
as implemented in the Quantum-ESPRESSO package \cite{QE}. The plane wave cutoff energy
for pure Ar is set to 476 eV and for all other systems to 544 eV. In the self-consistent
calculation of pure Ar a Monkhorst-Pack $32\times 32\times 32$ k-mesh is
employed, whereas for all other systems a $32\times 32\times 1$ k-mesh is used.
To achieve a high resolution, dense $64\times 64 \times 64$ and
$64\times 64\times 1$ k-meshes are used for calculating the density
of states (DOS). An energy convergence of $10^{-5}$ Ry and a force convergence of
$10^{-4}$ Ry/Bohr are achieved. Calculations are performed with and without
spin-orbit coupling (SOC) and with and without van der Waals (vdW) interaction \cite{london}.
In the zoomed band structures shown in the following we use
$\delta K=(0.002;0.002;0)$ with $K=(1/3;1/3;0)$. We consider a slab geometry with silicene
on one side of an Ar(111) slab, which inherits hexagonal symmetry from the two subsystems.

\section{Results and Discussion}

The optimized lattice parameter of solid Ar is $5.36$ {\AA}, which
leads to an Ar-Ar distance of $5.36/\sqrt{2}$ \AA $=3.79$ {\AA}. On the other hand,
for freestanding silicene we obtain $3.86$ {\AA} in agreement with
Ref.\ \cite{sili}. For the combined argon-silicene (ArSi) system we
set the lattice parameter to $3.79$ {\AA}, i.e., silicene is subject
to a lattice mismatch of 1.9\%. Figure \ref{fig:struct} shows for the ArSi system
a buckling of 0.53 {\AA} (distance between the bottom and top atomic layers),
which is slightly higher than predicted for freestanding
silicene (0.46 {\AA}) in Ref.\ \cite{sili1}. We note that an
artificially planarized structure with the same Si-Si bond length
as the ground state buckled structure is only 32 meV per Si atom higher in energy. The
Ar(111) substrate consists of six Ar layers. The top layer is
arranged such that two thirds of the atoms are located below Si
atoms and the last third is located below the center of a Si
hexagon, which turns out to minimize the energy.
After structural optimization without vdW interaction, the Si-Si
bond length is 2.27 {\AA} with a bond angle of $115^{\circ}$, in
agreement with Refs.\ \cite{sili1,sili2,sili3}. We obtain a distance
of $4.3$ {\AA} between the top Ar layer and the silicene, whereas
the interlayer spacing in the substrate amounts to 3.6 {\AA}. Taking into
account the vdW interaction, we obtain the same buckling but reduced
interlayer spacings of $3.4$ \AA\ and $3.1$ {\AA}, respectively.

\begin{figure}[t]
\begin{center}
\includegraphics[width=0.4\textwidth,clip=true]{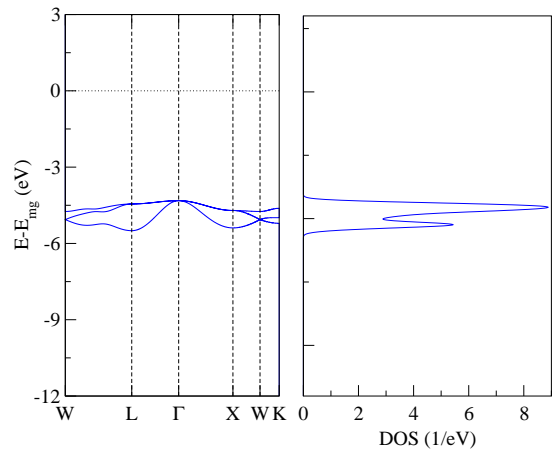}
\end{center}
\caption{\label{fig:ar1} Band structure and DOS of solid Ar (mg: midgap).}
\end{figure}

\begin{figure}[t]
\begin{center}
\includegraphics[width=0.45\textwidth,clip=true]{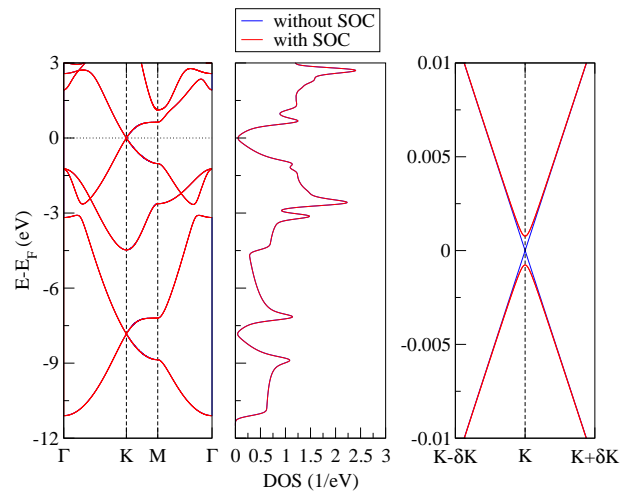}
\end{center} \caption{\label{fig:silicene} Band structure and DOS of silicene without vdW
interaction (no band gap without SOC and $E_{g}=2$ meV with SOC).}
\end{figure}

\begin{figure}[t]
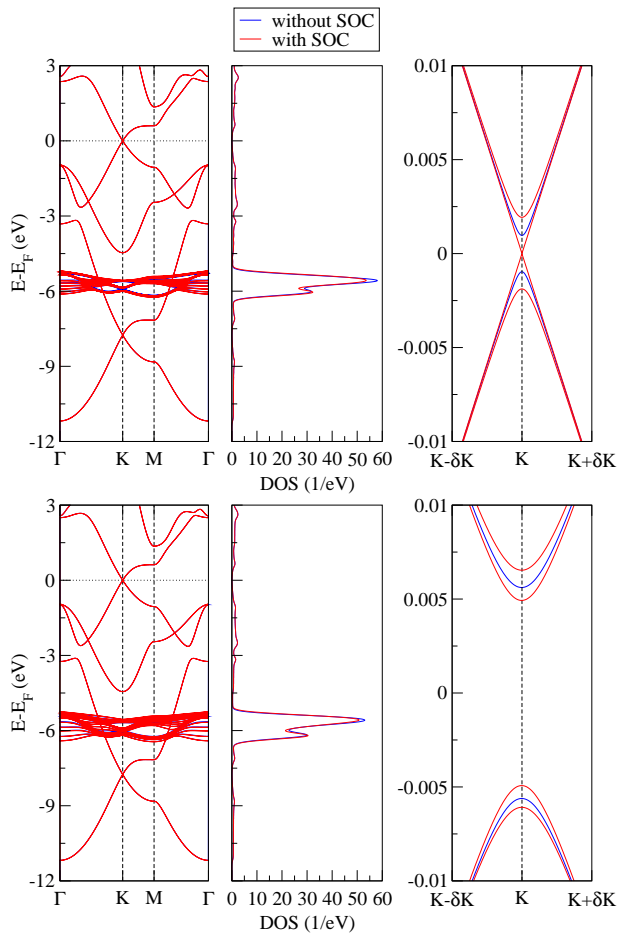

\begin{center}
\includegraphics[width=0.45\textwidth,clip=true]{arsi.eps}\\
\includegraphics[width=0.45\textwidth,clip=true]{varsi_band.eps}
\end{center}
\caption{\label{fig:arsi}Band structure and DOS of the ArSi system. Top: Without vdW
interaction ($E_{g}=2$ meV without SOC and no band gap with SOC).
Bottom: Including vdW interaction ($E_{g}=11$ meV without SOC and
$E_{g}=10$ meV with SOC).} 
\end{figure}

The band structure and DOS of solid Ar are shown in Fig.\ \ref{fig:ar1} and those of
freestanding silicene in Fig.\ \ref{fig:silicene}. Without SOC
we obtain for silicene the characteristic linear dispersion of the
$\pi$ and $\pi^*$ bands around the K point \cite{2dbands}, reflecting
massless Dirac fermions. The inclusion of SOC opens a band gap of 2 meV, which
is small, but much larger than in the case of graphene (due to the stronger SOC) and
agrees with Refs.\ \cite{sili3,si-hbn2}. When we turn on the vdW interaction we obtain
virtually identical results to Fig.\ \ref{fig:silicene} both without and with SOC.

For the combined ArSi system the band structure and DOS are
presented in Fig.\ \ref{fig:arsi}. Without vdW interaction and
without SOC we obtain again a band gap of 2 meV, indicating
minor influence of the substrate despite the fact that the interlayer
spacing between Ar and Si is large. SOC splits the bands near the K
point, one forming a perfect Dirac cone and one showing a band gap
of 2 meV. We note that the band gap of silicene can be
tuned by applying an external electric field, because the insulating Ar does not screen
electric fields. The binding energy $(E_{\rm ArSi}-E_{\rm Ar}-E_{\rm Si})/2$ per Si atom
between silicene and the substrate amounts to $-3$ meV, both without and with SOC.

Switching on the vdW interaction enhances the band gap to 11 meV
without and 10 meV with SOC. This is much larger than in graphene
but for many potential device applications still too small. For example, for
metal-oxide-semiconductor field-effect transistors a sizeable band gap
is required for a good on-off ratio and low power
dissipation \cite{device}. If SOC is included, both split bands show
energy gaps, where the splitting at the K point is of similar
magnitude as seen in the top part of Fig.\ \ref{fig:arsi}. The
larger band gap reflects the relevance of the vdW interaction in the
hybrid system, in contrast to freestanding silicene, while SOC
splits the bands near the Fermi level. The binding energy per Si
atom accordingly is enhanced to $-32$ meV, which, however, is
still very small. In particular, it is much less than reported for
silicene on inert hexagonal boron nitride \cite{si-hbn2}. Due to the
weak interaction, we conclude that solid Ar(111) will support quasi-freestanding silicene.

\section{Conclusion}

In conclusion, we have discussed the structure and electronic properties of silicene on
solid Ar(111). It turns out to be critical to take into account the vdW interaction to
obtain realistic results. We have shown that the Dirac cone of freestanding silicene
remains intact on Ar(111), which points to a weak interaction with the substrate. In fact,
we obtain for the binding energy a small value of $-32$ meV per Si atom, indicating a
quasi-freestanding nature of silicene on Ar(111). Any other substrate employed so far has
resulted in fundamental perturbations of the Dirac states, which is not the case on solid
Ar according to our simulations. It is likely that separation of silicene from this substrate
is possible.

\begin{acknowledgments}
Fruitful discussions with T.\ P.\ Kaloni are gratefully acknowledged.
We thank the subeditor for bringing to our attention the work of Weaver and coworkers
on noble gas substrates.
\end{acknowledgments}

\end{document}